\documentclass[prx,aps,twocolumn,superscriptaddress,longbibliography]{revtex4-1}%
\usepackage{graphicx}
\usepackage{dcolumn}
\usepackage{bm}
\usepackage{amssymb}
\usepackage{amsmath}
\usepackage{amsfonts}

\providecommand{\U}[1]{\protect\rule{.1in}{.1in}}
\usepackage{hyperref}
\hypersetup{
colorlinks,
citecolor=blue,
filecolor=blue,
linkcolor=blue,
urlcolor=blue
}

\begin{document}

\title{Weak Localization and Antilocalization in Topological Materials with Impurity Spin-Orbit Interactions}
\author{Weizhe Edward Liu} 
\affiliation{School of Physics and Australian Research Council Centre of Excellence in Low-Energy Electronics Technologies, UNSW Node, The University of New South Wales, Sydney 2052, Australia}
\author{Ewelina M. Hankiewicz}
\affiliation{Institute for Theoretical Physics and Astrophysics, W\"urzburg University, Am Hubland, 97074 W\"urzburg, Germany}
\author{Dimitrie Culcer}
\affiliation{School of Physics and Australian Research Council Centre of Excellence in Low-Energy Electronics Technologies, UNSW Node, The University of New South Wales, Sydney 2052, Australia} 

\begin{abstract}
Topological materials have attracted considerable experimental and theoretical attention. They exhibit strong spin-orbit coupling both in the band structure (intrinsic) and in the impurity potentials (extrinsic), although the latter is often neglected. In this work we discuss weak localization and antilocalization of massless Dirac fermions in topological insulators and massive Dirac fermions in Weyl semimetal thin films taking into account both intrinsic and extrinsic spin-orbit interactions. The physics is governed by the complex interplay of the chiral spin texture, quasiparticle mass, and scalar and spin-orbit scattering. We demonstrate that terms \textit{linear} in the extrinsic spin-orbit scattering are generally present in the Bloch and momentum relaxation times in all topological materials, and the correction to the diffusion constant is linear in the strength of the extrinsic spin-orbit. In topological insulators, which have zero quasiparticle mass, the terms linear in the impurity spin-orbit coupling lead to an observable density dependence in the weak antilocalization correction. They produce substantial qualitative modifications to the magnetoconductivity, differing greatly from the conventional Hikami-Larkin-Nagaoka formula traditionally used in experimental fits, which predicts a crossover from weak localization to antilocalization as a function of the extrinsic spin-orbit strength. In contrast, our analysis reveals that topological insulators \textit{always} exhibit weak antilocalization. In Weyl semimetal thin films having intermediate to large values of the quasiparticle mass, we show that extrinsic spin-orbit scattering strongly affects the boundary of the weak localization to antilocalization transition. We produce a complete phase diagram for this transition as a function of the mass and spin-orbit scattering strength. Throughout the paper we discuss implications for experimental work, and at the end we provide a brief comparison with transition metal dichalcogenides.
\end{abstract}
\date{\today}
\maketitle

\section{Introduction}

Topological materials such as topological insulators \cite{Hasan_TI_RMP10}, transition metal dichalcogenides \cite{Dixiao_MoS2_2012} and Weyl and Dirac semimetals \cite{Burkov_Weyl_TI, Fuhrer_Na3Bi_2016, Fuhrer_NL_Dirac_semi} have opened a new and active branch of condensed matter physics with considerable potential for spin electronics, thermoelectricity, magnetoelectronics and topological quantum computing \cite{SDS_TQC_RMP08}. The linear dispersion characterizing topological semimetals has attracted enormous attention both in experiment \cite{Wang_Bi2Te3_Ctrl_AM11, Benjamin_nature_2011,He11prl,Ren-PRB-2012, Kim_sur_2012, Jinsong-Zhang-Yayu-Wang-2013-science, Lucas_nl_tran_2014, Jianshi_nl_spin_polarized_2014, Kozlov_prl_2014, KimFuhrer_NC2013, Cacho_prl_2015, ZhangJinsong_prb_2015, Hellerstedt_APL_2014,Kastl_surface_tran_nature_2015, E_control_SOT_TI, Yoshimi15natcommun2} and in theory \cite{Hwang_Gfn_Screening_PRB07, JungMacDonald_graphene_PRB2011, Durst_2015, Haizhou_WL_2014, Zhang_TI_1stPrinc_NJP10, Shi_Rappe_nl_2016, Culcer_TI_AHE_PRB11,Adam_2D_Tran_prb_2012,Yoshida-PRB-2012,LiQiuzi_2013_prb,Weizhe_TITF_2014_prb,Lu14prl,Das_prb_2015}. Whereas the field is already vast, the focus of this paper will be primarily on topological insulators and Weyl and Dirac semimetals. In these materials, in which the band structure spin-orbit interactions are exceedingly strong, spin-orbit coupling in the impurity scattering potentials is also expected to be sizable, although its effect has been frequently neglected.
 
Topological insulators (TIs) are bulk insulators exhibiting conducting surface states with a chiral spin texture \citep{KaneMele_QSHE_PRL05, Hasan_TI_RMP10, Qi_TI_RMP_10, Moore_TRI_TI_Invariants_PRB07, Ando-TI, Culcer_TI_Int_PRB11, Culcer_TI_PhysE12, Tkachov_TI_Review_PSS13,Yong-qing_FOP_2012, Moore-nature}, in which the surface carriers are massless Dirac fermions described by a Dirac Hamiltonian \cite{Fu_3DTI_PRL07,Chen:2009}. Much of the excitement of recent years has been driven by the laboratory demonstration of magnetic topological insulators \citep{Hor_DopedTI_FM_PRB10,Jinsong-Zhang-Yayu-Wang-2013-science,Collins-McIntyre_Cr_Bi2Se3_2014, Wang15prl, Yoshimi15natcommun, Li15prl, Kou15natcommun, Katmis16nature, Jiang16aipadv, Fan16natnanotechnol, He17natmater}, in which the Dirac fermions have a finite mass and Berry curvature \cite{Dixiao_MoS2_2012,LuZhao_TITF_transport_PRL2013, Valley-Hall_graphene_prl_2007, Valley_prb_Niu, Bi-graphene_massive_prl}, which leads to the anomalous Hall effect \citep{Culcer_TI_AHE_PRB11, Yu_TI_QuantAHE_Science10, JiangQiao_TIF_QAHE_PRB2012, Chang_TIF_ferromagnetism_AHE_AM2013, Chang_QAHE_exper_Science2013}. Under certain circumstances this can be quantized and dissipationless \cite{Yu_TI_QuantAHE_Science10, FangZhongAHEReview2015, Liu16annurevcondensmatterphys}, as experiment has confirmed \cite{Chang_QAHE_exper_Science2013, Precise-QAHE-CZChang, D.Reilly_USYD, Kandala15natcommun, Chang15prl, Chang16jphyscondensmatter}, stimulating an intense search for device applications. Hybrid junctions between topological insulators and superconductors \citep{Sochnikov_TISC_nano_2013,Jin-Feng_TISC_2014} are expected to host topological superconductivity and Majorana fermions \citep{SDS_TQC_RMP08,Qiang-Hua_TISC_2014_SciRep}.

Weyl fermion semimetals are three-dimensional topological states of matter \cite{Potter_NC2014, Burkov_NM16, PhysRevX.6.021042, Yang_SPIN2016, Zhang_TI_NC14, Peng_NC2015,Li_CPB2013, PhysRevB.79.165422,Liu201672,Liu17prb, PhysRevB.82.155457, Oh153, Xu294,Song-Bo_NJP16,Lau_arxiv17}, in which the conduction and valence bands touch linearly at an even number of nodes, which appear in pairs with opposite chirality. First-principles theoretical studies as well as angle-resolved photo-emission spectroscopy experiments have confirmed the existence of chiral massless Weyl fermions \cite{RevModPhys.82.1959} in topological Dirac semimetals \cite{PhysRevB.85.195320, PhysRevB.88.125427, PhysRevLett.108.140405, Liu864,Xiong413,LiuNM,NeupaneNC,YiSR,PhysRevLett.113.027603,PhysRevLett.117.136602, Kushwaha_APLM2015, 0295-5075-114-2-27002, Haizhou_NC16, PhysRevLett.115.076601,Xiao_SR2015,Zhao_SR2016}, as well as in type-I \cite{PhysRevX.5.011029,TaAs_prediction,Xu613,PhysRevX.5.031013,doi:10.1063/1.4940924} and type-II \cite{TypeII_Weyl_intro,arXiv:1507.04847,Haizhou_TypeII,PhysRevB.93.121108} Weyl semimetals (WSM). In an ultrathin film of topological Weyl semimetal (WSM), the out-of-plane component of the momentum is quantized giving rise to a mass and an accompanying energy gap \cite{Lu10prb,PhysRevLett.106.136802}. Remarkably, the quantum spin and anomalous Hall effects may be observed in a single-compound device, a fact that has stimulated a considerable amount of recent interest in quantum transport in thin films of these topological semimetals.

Quantum transport at low temperatures is dominated by weak localization (WL) or antilocalization (WAL) effects \cite{PhysRevLett.113.247203, PhysRevB.92.045203, PhysRevX.5.031023, PhysRevB.83.165440, PhysRevB.88.081407, PhysRevB.91.245157, PhysRevB.92.075205, Haizhou_CPB16, Li2017, arXiv:1610.01413}. These corrections to the conductivity are noticeable when the quasiparticle mean free path is much shorter than the phase coherence length \cite{RevModPhys.80.1355} and arise as a result of the quantum interference between closed, time-reversed loops that circle regions in which one or more impurities are present \cite{RevModPhys.57.287}. Since the interference effects leading to WL/WAL disappear in weak external magnetic fields these corrections can typically be identified straightforwardly in an experiment, and are frequently used to characterize samples, in particular transport in novel materials. They provide valuable information about the system \cite{KimFuhrer_NC2013}, such as symmetries of the system, the phase coherence length \cite{Ewelina_PRL2014,Lu14prl}, and the quasiparticle mass \cite{Lu11prl}. Topological semimetals provide a new platform to understand weak localization and antilocalization behavior in generic 2D Dirac fermion systems in which the mass may be taken as a parameter \cite{Imura09prb, He11prl, Tkachov11prb, Lu11prl, PhysRevB.93.205306, PhysRevLett.89.046801, Shan12prb}. In the seminal work by Hikami, Larkin, and Nagaoka (HLN; \cite{Hikami80ptp}) for conventional electrons with a parabolic dispersion $\varepsilon_p = p^2/2m$, weak localization and antilocalization effects are classified according to the orthogonal, symplectic, and unitary symmetry classes, corresponding to scalar, spin-orbit, and magnetic impurities respectively. In topological insulators one expects weak antilocalization (WAL) in the presence of scalar disorder \cite{McCann:2006,Tkachov11prb,Adroguer:2012}. The WAL correction can be affected by the interaction of the surface states with the residual bulk states \cite{Garate:2012}, the thickness of the film \cite{Shan12prb}, or electron-electron interactions \cite{Koenig:2013,Lu14prl} changing its sign and turning it into weak localization (WL). In Dirac/Weyl semimetals the sign of the correction depends on the quasiparticle mass: at small mass it is identical to topological insulators, at large mass it is similar to ordinary massive fermions.

\begin{figure}[h]
\begin{center}
\includegraphics[width=0.5\columnwidth]{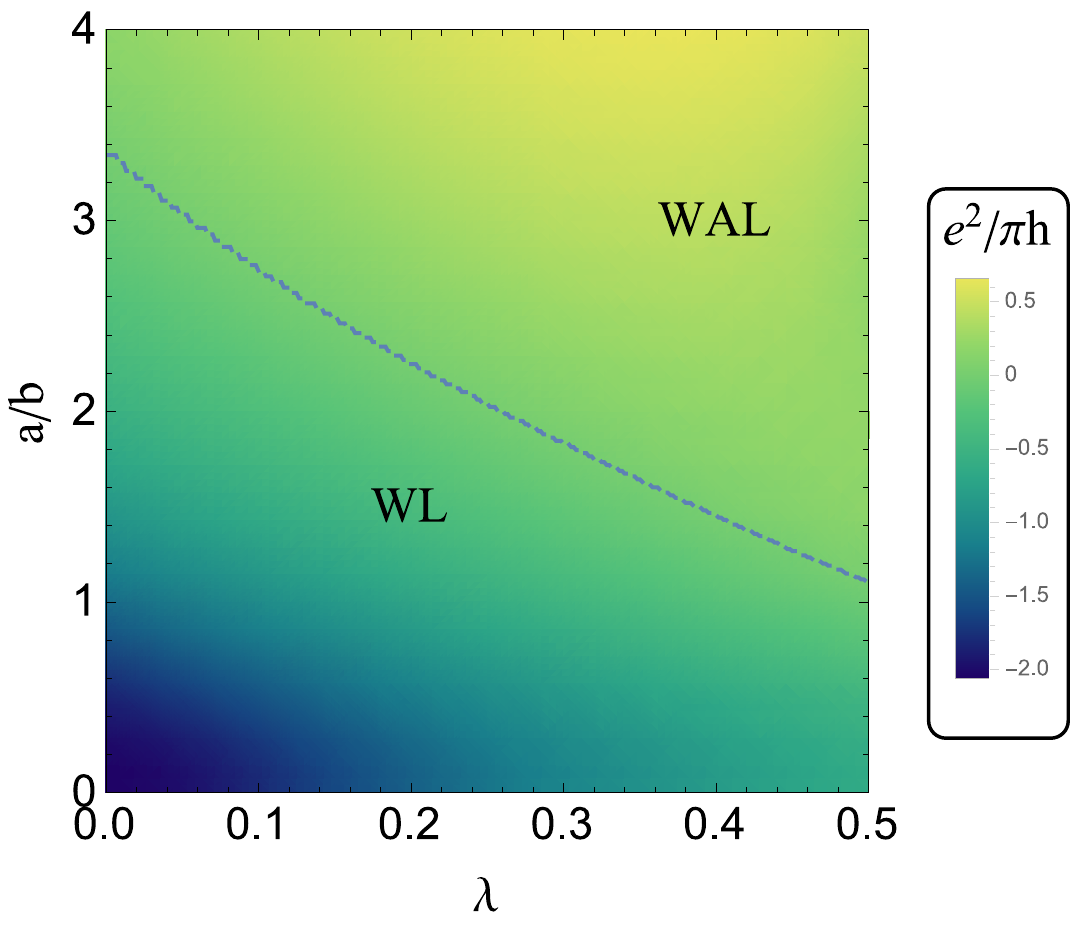}
\caption{Phase diagram representing the evolution of the quantum correction to the conductivity $\sigma^{\text{qi}}(0)$ at zero magnetic field for massless and massive Dirac fermions. The parameters $a$, $b$ and $\lambda$ are defined in Sec.~\ref{ModelFormalism}. The extrinsic spin-orbit scattering strength is quantified by $\lambda$, while the ratio $a/b$ represents the ratio of the spin-orbit energy to the quasiparticle mass, evaluated at the Fermi energy. The conductivity is expressed in units of $e^{2} / h$ with the color bar on the right. The blue dashed line separates the WAL and WL regimes. The numerical parameters here are the same as those used in Figure~\ref{eta1_Plot} below.}
\label{WALWLphase}
\end{center}
\end{figure}

In this article we will discuss weak localization and antilocalization of 2D fermions in topological semimetals while taking into account the strong spin-orbit coupling in the scattering potentials. We focus on two prototype systems: (i) topological insulators as representing massless Dirac fermions and (ii) Dirac/Weyl semimetal thin films as representing massive Dirac fermions. In order to capture the effect of strong spin-orbit scattering correctly it is necessary to treat scalar and spin-dependent scattering on the same footing, which requires one to retain the matrix structure of all the Green's functions and impurity potentials \cite{Culcer_TI_AHE_PRB11}. We will demonstrate that a \textit{linear} term is present in the strength of the extrinsic spin-orbit scattering potential, which has a strong angular dependence. Because of the winding of the spin around the Fermi surface, the Dirac nature of the surface states breaks the mirror symmetry around the $xy$-plane (the disorder potential no longer commutes with the kinetic Hamiltonian), and thus allows for a correction to physical quantities, such as the classical conductivity and the diffusion constant, which is \textit{linear} in the strength of the disorder spin-orbit coupling as opposed to the quadratic dependence observed for a parabolic dispersion. This term appears in the Bloch lifetime of the quasiparticles, the transport relaxation time, the spin relaxation time, and the Cooperon, and gives rise to a non-trivial density dependence of the quantum correction to the conductivity, which may be observable when the quasiparticle mass is very small, namely in topological insulators. They key point part of our analysis is provided by Eqs.~(\ref{zerofieldcondu})-(\ref{magnetoconduc}) below, which are completely general and can be used to fit WL/WAL experiments on both massless and massive Dirac fermions. Another important aspect is the fact that the extrinsic spin-orbit scattering suppresses the weak localization channel for massive fermions, and therefore has a strong qualitative and quantitative effect on the phase diagram of the weak localization to weak antilocalization transition, as seen in Fig.~\ref{WALWLphase}. The resulting weak localization/antilocalization behavior is consistent with the universality classes in the massless and massive limits respectively and with the symmetries of the system under chirality reversal. 

Importantly, the HLN formula commonly used to fit magnetoconductance experiments on Dirac fermions \cite{Checkelsky:2011,Zhang:2013,Lang:2013,KimFuhrer_NC2013,Bao:2013,Bardarson:2013, Akiyama_SnTe_2014} accounts neither for the Dirac nature of the electron states nor for their dimensionality. Whereas for electrons with parabolic dispersion a crossover from WL to WAL (no correction) is observed for 3DEGs (2DEGs), the correct formula for massless Dirac fermions always predicts WAL \textit{regardless} of the strength of the impurity spin-orbit coupling. The two formulas only converge at large strengths of the extrinsic spin-orbit scattering (when both equations in fact lose their validity). Moreover, the HLN formula gives the wrong value for the diffusion constant when the impurity spin-orbit coupling vanishes: it does not capture the fact that the absence of backscattering for massless Dirac fermions doubles the diffusion constant as compared to conventional electrons. Similar qualitative observations apply to massive Dirac fermions. 

This paper is organized as follows. In Sec.~\ref{ModelFormalism}, we describe a generic band structure model and theoretical transport formalism applicable to topological semimetals. Section~\ref{sec:TI} is devoted to quantum transport of massless fermions in topological insulators, discussing the conductivity dependence as a function of the external magnetic field and carrier density. In Sec.~\ref{Results} we discuss quantum transport of massive fermions in Weyl/Dirac semimetal thin films. In Sec.~\ref{Conclusions}, we summarize our conclusions briefly and discuss possible future research directions.

\section{General Model}
\label{ModelFormalism}

The systems discussed in this paper are described by the generic band Hamiltonian 
\begin{equation}\label{HGen}
H_{0\bm{k}} = \frac{\hbar}{2} \, {\bm \sigma} \cdot {\bm \Omega}_{\bm k}
\end{equation}
where ${\bm \sigma}$ is the vector of Pauli spin matrices and ${\bm \Omega}_{\bm k}$ plays the role of a wave-vector dependent effective magnetic field. For topological insulators ${\bm \Omega}_{\bm k} = (Ak_y, -Ak_x, 0)$ while for Weyl/Dirac semimetal thin films ${\bm \Omega}_{\bm k} = (Ak_x, Ak_y, M)$, where $A$ and $M$ are material/structure-specific constants. The eigenvalues of $H_{0\bm{k}}$ are $\displaystyle \varepsilon^{\pm}_{\bm{k}} = \pm \sqrt{A^{2} k^{2} + M^{2}} \equiv \pm \varepsilon_{k}$, where $+$ and $-$ indicate the conduction and valence bands, respectively, and $M = 0$ for TIs. We take the Fermi level to be located in the conduction band, and $ \varepsilon \sim \varepsilon_{\text{F}} = \sqrt{A^{2} k_{\text{F}}^{2} + M^{2}}$ at low temperatures, where $k_{\text{F}} $ is the Fermi wavevector. We abbreviate $a = A k_{\text{F}} / \varepsilon_{\text{F}}$ and $b = M / \varepsilon_{\text{F}}$ where $a^{2} + b^{2} = 1$. For our theory to be applicable we require $\varepsilon_F \tau_{tr}/\hbar > 1$, where $\tau_{tr}$ is the momentum relaxation time derived below. The $>$ sign means the Fermi wave vector is greater than the inverse of the mean free path, but not \textit{much} greater, so that disorder effects are observable. Conventionally, when one restricts one's attention to the Born approximation the requirement is $\varepsilon_F \tau_{tr}/\hbar \gg 1$, yet in this case weak localization/antilocalization corrections tend to be negligible.

The disorder potential consists of uncorrelated short-range impurities. The total impurity potential is $V(\bm{r}) = \sum_{I} U (\bm{r} - \bm{R}_{I})$, where $\sum_{I}$ is a summation over all impurities and $\bm{R}_{I}$ is the impurity coordinate. The matrix elements of a single impurity potential in reciprocal space, including the effect of extrinsic spin-orbit-coupling, take the form
\begin{equation}\label{SOIP}
U_{\bm{k}\bm{k}^{\prime}} = \mathcal{U}_{\bm{k}\bm{k}^{\prime}} \left( \openone + i \lambda \sigma_{z} \sin \gamma \right),
\quad \gamma =  \theta^{\prime} - \theta,
\end{equation}
where $\mathcal{U}_{\bm{k} \bm{k}^{\prime}} \equiv \mathcal{U}$ is the reciprocal-space matrix element of a short-range impurity potential, $\theta (\theta^{\prime})$ is the polar angle of the vector $\bm{k} (\bm{k}^{\prime})$, and $\openone$ represents the $2\times2$ identity matrix. It is assumed that $\lambda < 1$ since this term is typically treated in perturbation theory. Our simplified notation masks the fact that $\lambda \propto k_{\text{F}}^{2} = 4 \pi n_{e}$ is a linear function of the electron density $n_e$, which experimentally can be tuned by changing the gate voltage or the temperature \cite{Shekhar_NP2015}. Notice that the effect of the spin-orbit coupling term in the scattering potential is to create a random, effective magnetic field, which points either into or out of the plane of the structure. \footnote{In analytical calculations of transport coefficients dealing with explicit impurity configurations is unwieldy. The customary solution to this is to perform an average over all the possible impurity configurations, whereupon the total impurity potential in reciprocal space $V_{\bm{k}\bm{k}^{\prime}}$ becomes \cite{PhysRevB.95.014204} $\overline{V^{\alpha\beta}_{\bm{k}\bm{k}^{\prime}}} = 0$ and $\overline{V_{\bm{k}\bm{k}^{\prime}}^{\alpha\beta} V_{\bm{k}_{1}\bm{k}_{1}^{\prime}}^{\eta\zeta}} = n_{i} U^{\alpha\beta}_{\bm{k}\bm{k}^{\prime}} U_{\bm{k}_{1}\bm{k}_{1}^{\prime}}^{\eta\zeta}$
where $ \alpha,$ $\beta,$ $\eta,$ and $\zeta$ are spin indices, and  $n_{i}$ representing the impurity concentration.}

\begin{figure}[h]
\begin{center}
\includegraphics[width=0.2\columnwidth]{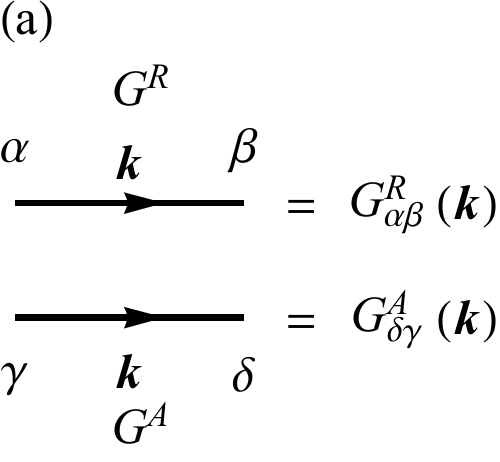} \quad
\includegraphics[width=0.3\columnwidth]{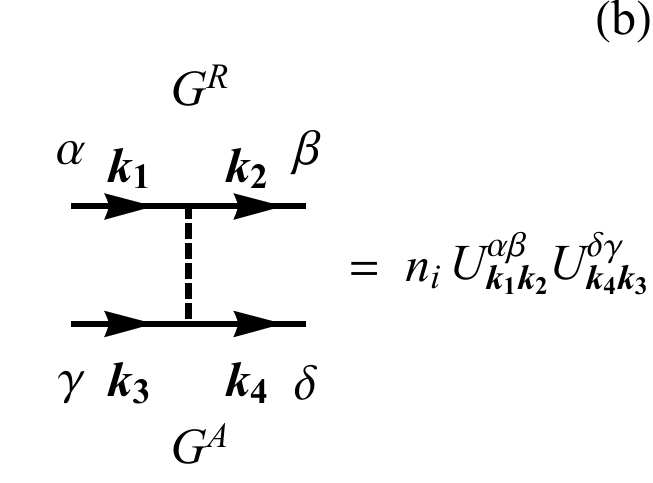} \\[-3ex]
\includegraphics[width=0.2\columnwidth]{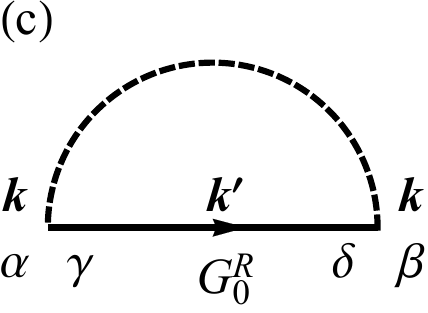} \quad \quad
\includegraphics[width=0.3\columnwidth]{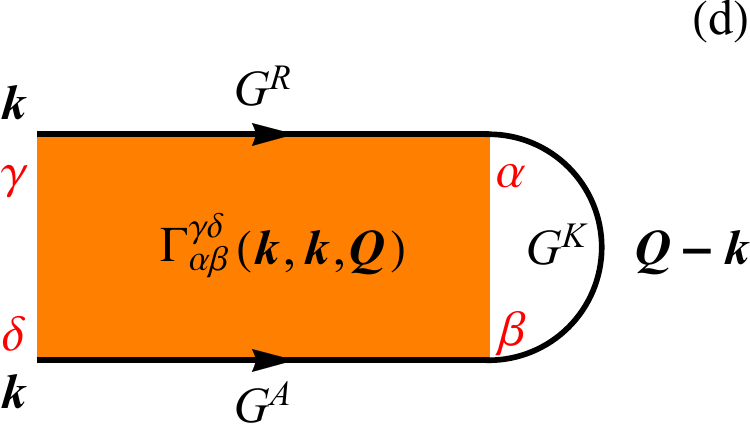} \\
\includegraphics[width=0.55\columnwidth]{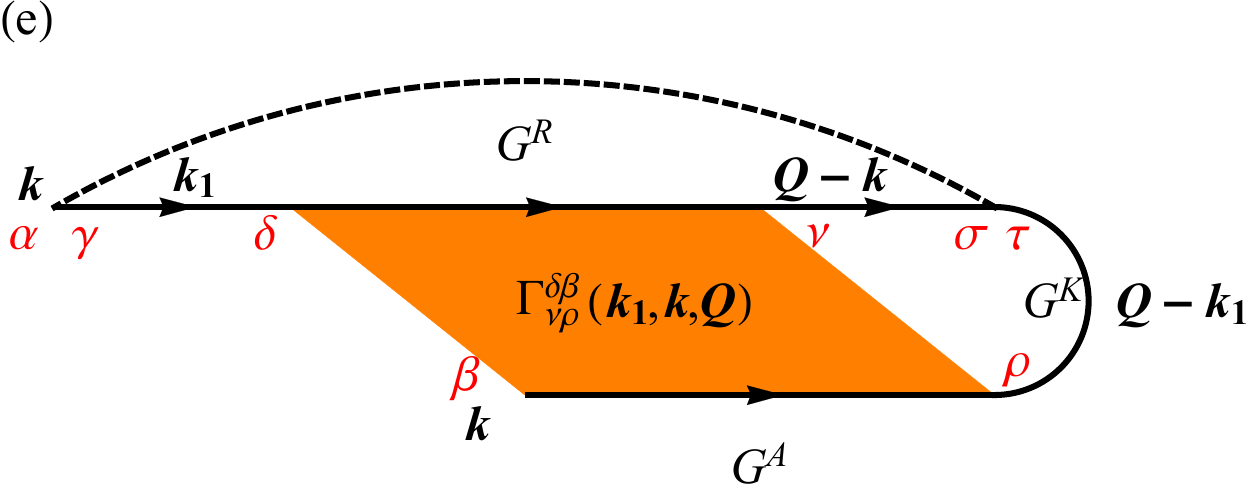} \\[-3ex]
\includegraphics[width=0.6\columnwidth]{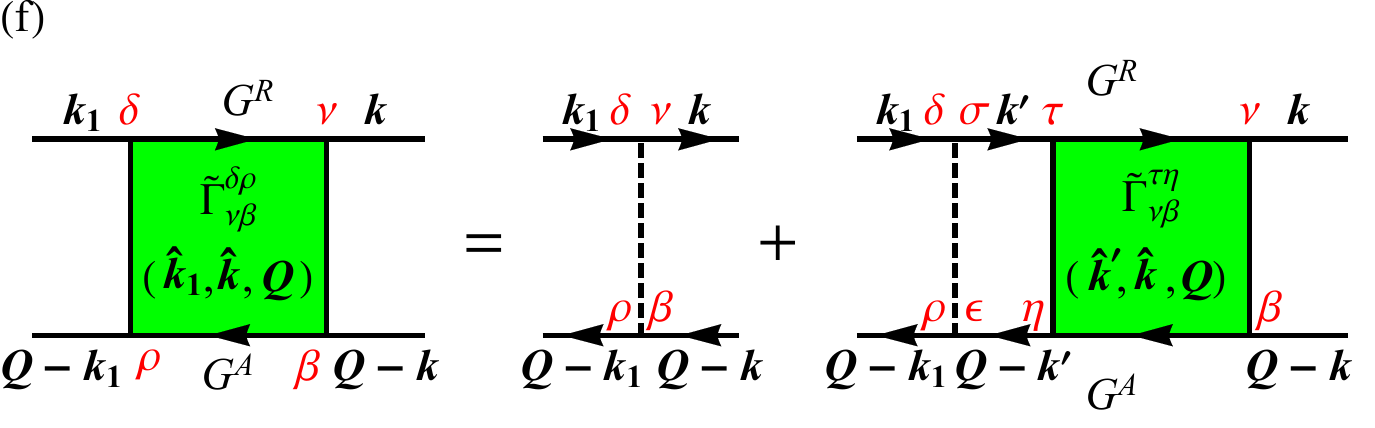}
\caption{The diagrams for the weak (anti-)localization conductivity of Dirac fermions.
(a) Definition of the Green's functions as arrowed solid lines in which Greek letters are spin indices.
(b) Definition of dashed lines: impurities lines expressed in both retarded and advanced cases.
(c) The retarded self-energy in the first-order Born approximation, where $G^{\text{R}}_{0}$ is the bare retarded Green's function. This contribution to the self energy represents the classical picture of electrons scattering off randomly distributed impurities. 
(d) and (e) are Keldysh self-energies $\big(\Sigma^{\text{K},\text{b}}_{\bm{k},\gamma\delta}$ and $\Sigma^{\text{K},\text{R}}_{\bm{k},\alpha\beta} \big)$ of maximally crossed diagrams in the bare and the retarded dressed cases, respectively, where $\Gamma$ is the Cooperon structure factor. Both of these represent contributions due to quantum interference in scattering processes: an electron travelling through a disordered region can backscatter and return to its starting point. The loop can be performed clockwise or anticlockwise, and in quantum mechanics the two trajectories interfere. Note that (e) vanishes in the absence of spin-orbit coupling. (f) The Bethe-Salpeter equation for the twisted Cooperon structure factor $\tilde{\Gamma}.$}
\label{All_Diagrams}
\end{center}
\end{figure}

Our discussion is based on the Keldysh \cite{Keldysh} transport formalism, whose central quantity is the Keldysh Green's function $G^{\text{K}}$. Its physical content is analogous to that of the density matrix $\rho$ in the quantum Liouville equation. In the Keldysh representation, the Keldysh Green's function $G^{\text{K}}$, together with the retarded and advanced Green's functions $G^{\text{R}}$ and $G^{\text{A}}$, form a $2\times2$ matrix:
\begin{equation}\label{GK}
\check{G} = \left(
\arraycolsep 1ex
\begin{array}{cc}
0 & G^{\text{A}} \\[1ex]
G^{\text{R}} & G^{\text{K}}
\end{array}
\right).
\end{equation}
The Green's function matrix $\check{G}$ is related to the \textit{contour-ordered} Green's function $\hat{G}$ through the equation $\hat{G} = R \, \check{G} \, R^{-1}$, with $R = (1+i\tau_{y})/\sqrt{2}$, where the contour takes into account time evolution in both directions and $\tau_y$ is a Pauli matrix in the $2 \times 2$ Keldysh space of Eq.~\ref{GK}. The Dyson equation for $\hat{G}$ may be written as
\begin{equation}
\hat{G} (1,2) = \hat{g} (1,2) + \int \text{d} 3 \int \text{d} 4  \, \hat{g} (1,3) \hat{\Sigma} (3,4) \hat{G} (4,2).
\end{equation}
where $\text{d} 3 = \text{d} \bm{r}_{3} \, \text{d} t_{3}$ etc. The self-energy matrix $\hat{\Sigma}$ is
\begin{equation}
\hat{\Sigma} = R \left(
\begin{array}{cc}
\Sigma^{\text{K}} & \Sigma^{\text{R}} \\
\Sigma^{\text{A}} & 0
\end{array}\right) R^{-1},
\end{equation}
where $\Sigma^{\text{K}},$ $\Sigma^{\text{R}},$ and $\Sigma^{\text{A}}$ are the Keldysh, retarded, and advanced self-energies. 

To illustrate the above, in Figure~\ref{All_Diagrams} the arrows represent Green's functions as defined in Figure~\ref{All_Diagrams}(a), and the dashed lines in Figure~\ref{All_Diagrams} correspond to the impurity scattering whose definitions in the retarded and the advanced cases are displayed in Fig~\ref{All_Diagrams}(b). After the average over impurity configurations, the retarded component of the self-energy in the Born approximation is shown in Figure~\ref{All_Diagrams}(c), while the Keldysh components of the self-energy leading to quantum-interference are in Figs.~\ref{All_Diagrams}(d) and \ref{All_Diagrams}(e) \cite{Vasko}. In Figure~\ref{All_Diagrams} only the spin indices are explicitly retained.

In the Born approximation the disorder-averaged retarded and advanced Green's functions \footnote{To get the bare Green's function $G_{0\bm{k}}$ simply remove the term containing $\tau$.} are
\begin{equation}
\hspace{-0.18cm} G^{\text{R}}_{\bm{k}} = \frac{\openone + H_{0\bm{k}} / \varepsilon_{k}}{2(\varepsilon - \varepsilon_{k} + \frac{i \hbar}{2\tau} )} = \big[ G^{\text{A}}_{\bm{k}} \big]^{*},
\end{equation}
where $^*$ denotes the Hermitian conjugate, $G^{\text{A}}_{\bm{k}}$ is the advanced Green's function, and $\tau$ is the elastic scattering time (Bloch lifetime)
\begin{equation}\label{tau}
\tau = \tau_{0} / \big[ \big( 1+ \lambda^{2}/2 \big) \big( 1 + b^{2} \big)
+ \lambda a^{2} \big],
\end{equation}
where $1/\tau_{0} = \pi n_{i} |\mathcal{U}|^{2} N_{\text{F}} /\hbar$. Note that we have retained the matrix structure of all Green's functions as well as that of the scattering potential, so that the elastic scattering time has corrections already at order $\lambda$-linear. This is unlike the case of non-relativistic electrons \cite{Hikami80ptp}: in the conventional Hikami-Larkin-Nagaoka approach the square of the extrinsic spin-orbit coupling terms is averaged over the Fermi surface. Here the mean-free time is no longer an even function of $\lambda$, since the winding of the spin around the Fermi surface for Dirac fermions defines unequivocally the direction of the $z$-axis. The transport time $\tau_\text{tr}$, in which the weight of small-angle scattering is reduced as compared to the Bloch lifetime, appears in the diffusion constant $D = \frac{v_\text{F}^2 \tau_\text{tr}}{2}$, making the diffusion constant dependent on the strength of the impurity spin-orbit coupling to \textit{linear} order. As this diffusion constant is a crucial parameter in weak antilocalization, we expect the behavior of Dirac fermions to be different from the usual HLN formula \cite{Hikami80ptp}.

In a constant, uniform external electric field $\bm{E}$, the quantum kinetic equation for $G^{\text{K}}$ is \cite{Rammer}
\begin{equation}\label{Kineticeq}
\partial_{t} G^{\text{K}}_{\varepsilon {\bm k}} + ( i / \hbar ) \big[ H_{0{\bm k}} , G^{\text{K}}_{\varepsilon {\bm k}}\big]
+ \mathcal{J}_{\varepsilon {\bm k}} = (e/ \hbar){\bm E} \cdot \partial_{\bm{k}} G^{\text{K}}_{\varepsilon {\bm k}},
\end{equation}
where $\partial_{\bm{k}} \equiv \partial / \partial {\bm k}$, the electron charge is $-e$, $\varepsilon$ is an intermediate energy variable that is integrated over (representing non-locality in time), and the general scattering term $\mathcal{J}_{\varepsilon{\bm k}}$ is given by
\begin{equation}\label{J_epsilon_k}
\mathcal{J}_{\varepsilon{\bm k}} \!= \! ( i / \hbar ) \big(\Sigma^{\text{R}}_{\varepsilon{\bm k}}
G^{\text{K}}_{\varepsilon{\bm k}} - G^{\text{K}}_{\varepsilon{\bm k}} \Sigma^{\text{A}}_{\varepsilon{\bm k}}
+ \Sigma^{\text{K}}_{\varepsilon{\bm k}} G^{\text{A}}_{\varepsilon{\bm k}}
- G^{\text{R}}_{\varepsilon{\bm k}} \Sigma^{\text{K}}_{\varepsilon{\bm k}} \big).
\end{equation}
We adopt the form $\displaystyle G^{\text{K}}_{\varepsilon {\bm k}} = \chi_{\bm k} (G^{\text{R}}_{\varepsilon {\bm k}}$ $- G^{\text{A}}_{{\varepsilon}{\bm k}} )$ with $\chi_{\bm k}$ a scalar, following the reasoning of Ref.~\cite{Schwab_TI_LokTnl_EPL11}. The Wigner transformation is applied on $\hat{G}(1,2)$ to find the single-particle Green's function $\hat{G}_{\varepsilon\bm{k}}$.

With the self-energy in the Born approximation as shown in Figure~\ref{All_Diagrams}(c), the scattering term becomes
$\mathcal{J}_{\text{Bn}} (\chi_{\bm{k}}) = - 2 \, i \pi \, g_{\bm{k}} \, \chi_{\bm{k}} / \tau_{\text{tr}}$, with
\begin{equation}\label{trsptime}
\tau_{\text{tr}} = 2 \tau_{0} / \big[1 + 3 b^{2} + 2 a^{2} \lambda +
\big( 5 + 3 b^{2}\big) \lambda^{2} /4 \big].
\end{equation}
Again, the corrections due to spin-orbit scattering appear at order $\lambda$-linear. The mean free path is $\ell_{e} = \sqrt{D \tau_{\text{tr}}}$ where $D = v_{\text{F}}^{2} \tau_{\text{tr}} / 2$ is the diffusion constant and $v_{\text{F}}$ the Fermi velocity. In the Born approximation the kinetic equation takes the form:
\begin{equation}
\mathcal{J}_{\text{Bn}} (\chi_{E\bm{k}}) = (e/ \hbar){\bm E} \cdot [\partial_{\bm{k}} G^{\text{K}}_{\varepsilon {\bm k}}]_\parallel,
\end{equation}
where the subscript $\parallel$ indicates the matrix component of the driving term that commutes with the band Hamiltonian $H_{0\bm{k}}$. To the leading order in the quantity $\tau_{\text{tr}}^{-1}$ the solution of Eq.~(\ref{Kineticeq}) is $\chi_{E\bm{k}} = [2 e \bm{E} \cdot \hat{\bm{k}} \, \tau_{\text{tr}} / \hbar] \delta (k - k_{\text{F}})$
where $\hat{\bm{k}}$ is the unit vector along $\bm{k}$, and the Drude conductivity takes the form
\begin{equation}
\sigma^{\text{Dr}}_{xx} = (e^{2}/h) (v_{\text{F}} \tau_{\text{tr}} / 2).
\end{equation}
Since this formula is general and well known, we shall not discuss it further. 

Quantum interference contributions to the conductivity are intimately connected to the phase information of the electron wave function. This is preserved over a distance referred to as the phase coherence length $\ell_{\phi} = \sqrt{D \tau_\phi}$, with $\tau_\phi$ the phase coherence time. Mechanisms giving a finite phase coherence time are associated with time-dependent perturbations such as phonons or electron-electron scattering. The observation of weak localization and antilocalization requires the mean free path $\ell_{e}$ to be much shorter than the phase coherence length $\ell_{\phi}$. Correspondingly, the phase coherence time $\tau_\phi$ is much longer than either the Bloch lifetime or the momentum relaxation time. Typically the calculation of the phase coherence time is extremely complex and controversial, and in almost all studies this quantity is typically extracted by fitting to experiment. 

The quantum-interference between two time-reversed closed trajectories will generate three different contributions
to the conductivity, whose Keldysh self-energies are (bare case) $\Sigma^{\text{K}}_{\text{b}},$
(Retarded dressed case) $\Sigma^{\text{K}}_{\text{R}},$
and (Advanced dressed case) $\Sigma^{\text{K}}_{\text{A}}.$
The Keldysh self-energies diagrams of $\Sigma^{\text{K}}_{\text{b}}$ and $\Sigma^{\text{K}}_{\text{R}}$
are shown in Figs.~\ref{All_Diagrams}(d) and \ref{All_Diagrams}(e), respectively.
In these diagrams, the maximally crossed diagrams
(Cooperon structure factor) $\Gamma$ is proportional to $1/|\bm{Q}|^{2},$
where $\bm{Q}$ is the sum of the incoming and the outgoing wavevector.
The divergence of $\Gamma_{\bm{Q}}$ at $|\bm{Q}| \to 0$ indicates the primary contribution into
the quantum-interference conductivity comes from the backscattering.
The quantum-interference self-energies ($\Sigma^{\text{K}}_{\text{b}},$ $\Sigma^{\text{K}}_{\text{R}}$
and $\Sigma^{\text{K}}_{\text{A}}$) will generate the quantum-interference scattering term
$\mathcal{J}_{\text{qi}} (\chi_{E\bm{k}})$ that is balanced by $\mathcal{J}_{\text{Bn}} (\chi_{\text{qi},\bm{k}})$ as follows:
\begin{equation}
\mathcal{J}_{\text{Bn}} (\chi_{\text{qi},\bm{k}}) = - \mathcal{J}_{\text{qi}} (\chi_{E\bm{k}}),
\end{equation}
where the right-hand side plays the role of a driving term. The term $\chi_{\text{qi},\bm{k}}$ found from this equation leads to the quantum-interference conductivity correction
\begin{equation}\label{sigmaWAL_Ini}
\sigma^{\text{qi}}_{xx} = - \frac{e^{2} v_{\text{F}}^{2} \tau \tau_{\text{tr}}^2 N_{\text{F}} \eta_{v}^{2}}{\hbar^{2}} \int \frac{\text{d}^2Q}{(2 \pi)^{2}} \, \big[ C^{\text{b}}_{\bm{Q}} + 2 C^{\text{R}}_{\bm{Q}} \big],
\end{equation}
where $ C^{\text{b}}_{\bm{Q}}$ and $ C^{\text{R}}_{\bm{Q}}$ are the bare Cooperon and the retarded-dressed Cooperon, respectively, both shown in Figure~\ref{All_Diagrams}.
Here $ C_{\bm{Q}} = \text{Re} \big[ \Sigma^{\text{K}}_{\bm{k}, \gamma\alpha}  g_{\bm{k}}^{\alpha \gamma}  \big]$ in general, and $\alpha$ and $\gamma$ are spin indices.
The Einstein summation rule over repeated indices is used throughout this work. Note that the advanced dressed Cooperon contributes the exactly same amount as its retarded dressed counterpart.

The zero magnetic-field conductivity can be obtained by integrating Eq.~(\ref{sigmaWAL_Ini}) over the magnitude of the momentum transfer $Q \equiv |\bm{Q}|$. Formally the integral diverges at low $Q$, and for this reason it must be regularized by cutting it off at a wave vector $1/\ell_\phi$, which is the inverse of the phase coherence length, which is generally by far the largest length scale in the problem. The upper limit of integration is typically taken as the inverse of the mean free path. This yields
\begin{equation}\label{zerofieldcondu}
\sigma^{\text{qi}}_{xx} (0) = \frac{e^{2}}{\pi h} \sum_{i=1,2,3} \alpha_{i} \ln
\bigg(\frac{1/\ell_{i}^{2} + 1/\ell_{\phi}^{2}}{1/\ell_{i}^{2} + 1/\ell_{e}^{2}}\bigg),
\end{equation}
where $i = 1, 2,$ and 3 are the singlet, the triplet-up, and the triplet-down channel indices, respectively, $\alpha_{i}$ is the weight of channel $i$ and $\ell_{i}$ is the effective mean free path for channel $i$. Both $\alpha_i$ and $\gamma_i$ contain contributions of first order in the extrinsic spin-orbit scattering strength. In the massless limit $(b = 0)$, $\alpha_{1} = -1/2$ and $\gamma_{1} = 0$, matching the findings of Ref.~\cite{Adroguer15prb}. In an out-of-plane magnetic field $B$ the function $\Delta \sigma (B) = \sigma^{\text{qi}}_{xx}(B) - \sigma^{\text{qi}}_{xx} (0)$ is described by the general expression \cite{Fuhrer_NL_Dirac_semi, Lu2016}
\begin{equation}\label{magnetoconduc}
\Delta \sigma (B) = \frac{e^{2}}{\pi h} \sum_{i =1,2,3} \alpha_{i}
\bigg( \Psi \bigg[ \frac{B_{\phi,i}}{ | B |} + \frac{1}{2} \bigg] - \ln \bigg[ \frac{B_{\phi,i}}{ | B | } \bigg] \bigg),
\end{equation}
where $B_{\phi,i} = (\hbar/4e) (1/\ell_{\phi}^{2} + 1/\ell_{i}^{2})$, and $\Psi$ is the digamma function. This equation can be used to fit magnetoconductance experiments on Dirac fermions, regardless of whether they are massless or massive. It accounts \textit{fully} for both linear and quadratic extrinsic spin-orbit coupling terms. 

\section{Topological insulators}
\label{sec:TI}

The Dirac Hamiltonian for the surface states of 3DTIs takes the form
\begin{equation}
H_{0{\bm k}} = A (\sigma_x k_y - \sigma_y k_x),
\label{eq:ham_kin}
\end{equation}
where $A$ is a material specific constant. The Hamiltonian~(\ref{eq:ham_kin}) preserves time-reversal symmetry (TRS), characteristic of the symplectic symmetry class. We have stated above that the Fermi energy is located in the conduction band, and for TIs we need to make the additional specification that it lies in the \textit{surface} conduction band, so that any possible bulk channels are not considered. This is justified by the fact that experiment has striven for a decade to overcome the bulk contribution to transport, and in a recent work it was reported that bulk carriers have been completely eliminated \cite{Ngabonziza_Bi2Se3TF_Tunable_AEM16}.

Since topological insulator samples are always films, a series of remarks are in order concerning the applicability of our theory. In an ultrathin film, namely thinner than 6 quintuple layers \cite{Linder_TI_size_PRB2009, Liu_TIF_PRB2010, LuShan_TITF_MassiveDirac_spinphys_PRB2010, Park_TITF_topoloprotect_PRL2010, Chang_TIF_ferromagnetism_AHE_NP2013, Hirahara_TITF_transition_PRB2010, Sakamoto_TITF_transition_PRB2010, Taskin_TITF_transport_PRL2012} tunneling is enabled between the surfaces. This gives a mass in the Dirac dispersion and, with time reversal symmetry preserved, the conduction and valence bands are each twofold degenerate. Quantum transport in such systems is extremely complex even in the Born approximation \cite{Weizhe_TITF_2014_prb}, with exciting predictions in the presence of carrier-carrier interactions, such as topological exciton condensation \cite{Seradjeh_TIF_excitoncondens_chargefract_PRL2009, Seradjeh_TITF_TEC_PRB2012, Efimkin_TIF_EHP_PRB2012, KimHankiewicz_TI_ExcitonSuperfluid_PRB2012} and quantum Hall superfluidity \cite{Tilahun_TIF_QHS_PRL2011}. They will not be of concern to us in this work, in which we will focus on relatively thick films, of approximately 10 quintuple layers and above. The physics of films is determined by the thickness $d$ and the Fermi wave vector $k_F$ \cite{Tilahun_TIF_QHS_PRL2011}. Consider a Bi$_2$Se$_3$ film as an example, with a dielectric constant $\kappa \approx 100$, grown on a semiconductor substrate with $\kappa_{s} \approx 11$. The condition for the film to be thick is $k_Fd \gg 1$. In this case the two surfaces are independent \cite{AndoFowler_2D_RMP1982}. For the top surface, where one side is exposed to air, $\kappa_{top} = (\kappa + 1)/2 \approx 50$. For the bottom surface $\kappa_{btm} = (\kappa + \kappa_{s})/2 \approx 55$. Both are independent of $d$.

\begin{figure}
\centering
\includegraphics[width= 0.50 \textwidth]{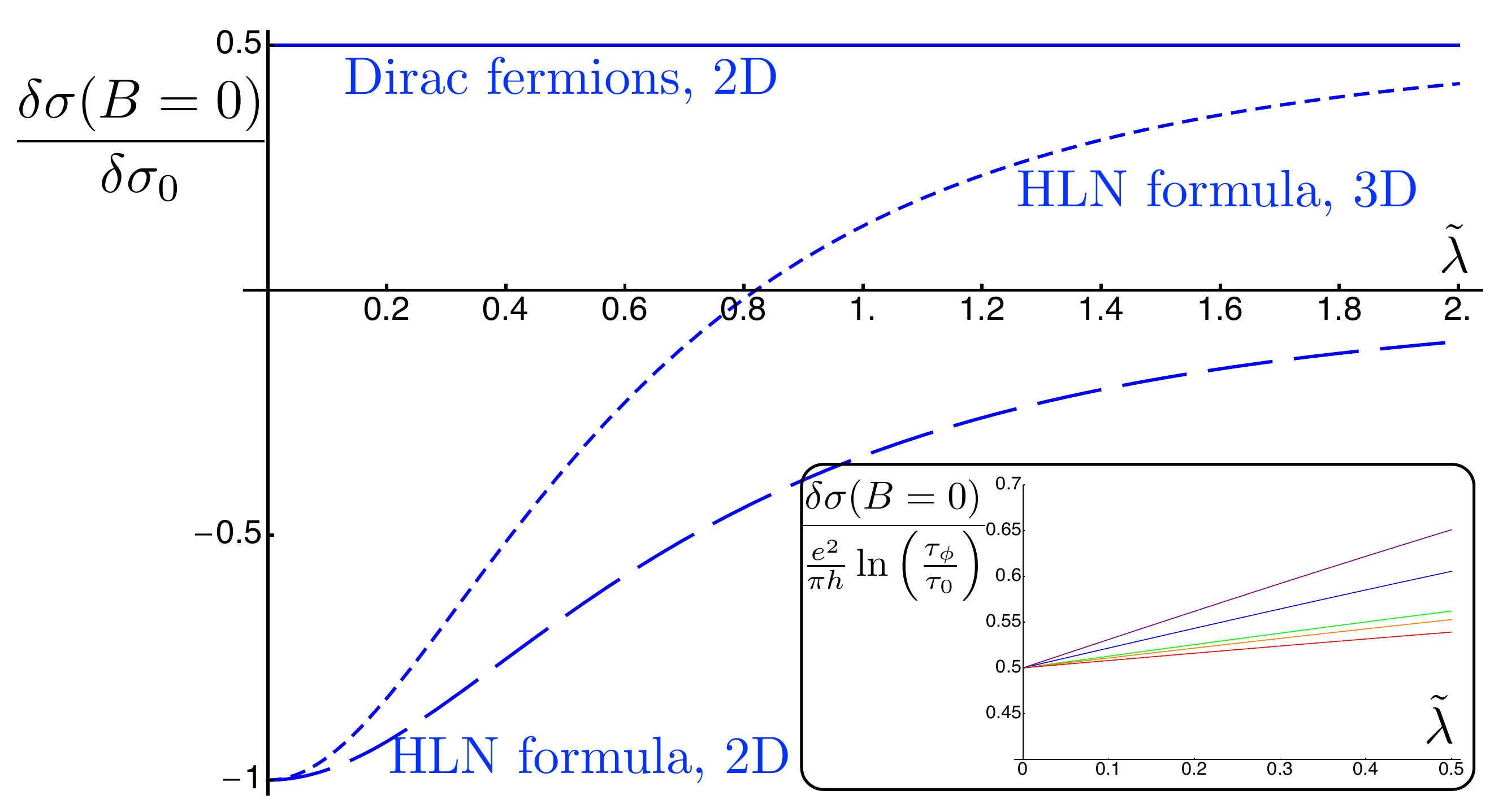}
\caption{Plot of the zero-field quantum-interference conductivity correction in terms of the spin-orbit scattering strength
$\lambda$ for massless Dirac fermions (solid line),  HLN formula for 2DEG (dashed line) and
3DEG (dotted line). $\delta \sigma_0 =  (e^2 / \pi h) \ln ( \tau_{\phi} / \tau )$.
In order to emphasize that the WAL conductivity of massless Dirac fermions is also dependent upon
the spin-orbit impurity strength,
the inset shows WAL in units independent of $\lambda$, i.e. $(e^2 / \pi h) \ln ( \tau_{\phi} / \tau_{0} ),$ 
where $\tau_0$ is the scattering time in the absence of spin-orbit impurities and
the ratio $\tau_\phi / \tau_0$ varies from 5 (purple), 10 (blue), 50 (green), 100 (orange) to 500 (red).
Adapted from Figure~3 in Ref.~\cite{Adroguer15prb}.}
\label{fig:Qcorrec_alpha}
\end{figure}

The WAL correction in the absence of a magnetic field takes the form:
\begin{equation}
\sigma_{xx}^{qi}(0) = \frac{e^2}{2 \pi h} \ln \left( \frac{\tau_{\phi}}{\tau} \right) \label{eq:delta1}.
\end{equation}
This expression should be compared with the formula for non-relativistic electrons
\begin{equation}
\arraycolsep 0.3ex
\begin{array}{rl}
\displaystyle \sigma_\text{xx, HLN}^\text{qi (2D)} (0) = & \displaystyle - \frac{e^2}{\pi h} \ln \left( \frac{1 + \lambda^2/2}{ \frac{\tau}{ \tau_{\phi}} + \lambda^2/2} \right) \\ [3ex]
\displaystyle \sigma_\text{xx, HLN}^\text{qi (3D)} (0)= & \displaystyle - \frac{e^2}{\pi h} \Bigg\{ \ln \frac{1 + \frac{2 \lambda^2}{3}}{ \frac{\tau}{ \tau_{\phi}} + \frac{2 \lambda^2}{3}}  - \frac{1}{2}  \ln \bigg[1+ \frac{8 \lambda^2}{9}\frac{\tau_{\phi}}{\tau}\bigg] \Bigg\}.
\end{array}
\end{equation}
These three formulas are plotted in Figure~\ref{fig:Qcorrec_alpha} as a function of $\lambda$. We have renormalized these corrections by $\delta \sigma_0 =  \frac{e^2}{\pi h} \ln \left( \frac{\tau_{\phi}}{\tau} \right)$, where $\tau$ depends on the model and is a function of $\lambda$. We have set $\frac{\tau_{\phi}}{\tau_0} = 10$ in agreement with what is measured experimentally\cite{Ewelina_PRL2014,Chiu:2013,Zhao:2013}. We observe that the Dirac fermions remain in the same symmetry class (symplectic, with WAL), whereas the HLN formula shows a crossover from the orthogonal symmetry class (WL) to either no correction in 2D, or WAL in 3D.

%------------------------------------------------------------
%------------------------------------------------------------

Due to the renormalization of the scattering time by spin-orbit impurities, one could also interpret Eq.~(\ref{eq:delta1}), as increasing with $\lambda$ if one renormalizes conductivity corrections by the scattering time in the absence of spin-orbit impurities $\tau_0$ (namely normalized by $\ln \tau_{\phi}/\tau_{0}$) : 
\begin{equation}
\sigma_{xx}^{qi}(0) = \frac{e^2}{2 \pi h}\left[ \ln \left( \frac{\tau_{\phi}}{\tau_{0}} \right) + \lambda + O(\lambda^2) \right].
\end{equation}
The inset to Figure~3 shows the linear dependence of the normalized correction to the conductivity as a function of $\lambda$ for massless Dirac fermions. As $\lambda = \lambda_0 k_\text{F}^2$, one can experimentally probe this linear dependence by varying the density using an electrostatic gate. 

%------------------------------------------------------------
\begin{figure}
\centering
\includegraphics[width= 0.45 \textwidth]{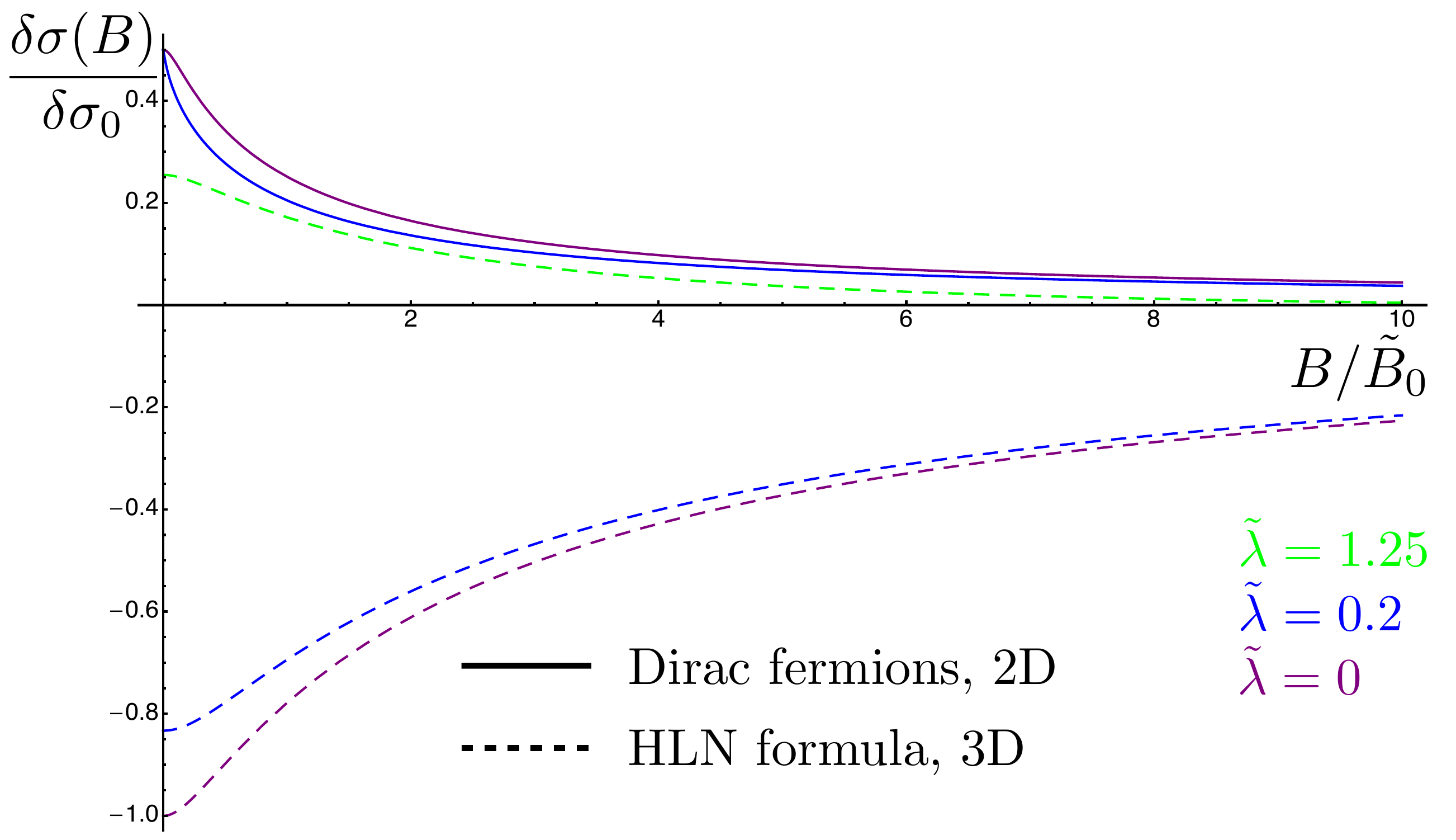}
\caption{The magnetic field dependence of the quantum correction to the conductivity for massless Dirac fermions (solid lines) and the 3D HLN formula (dashed lines)
at different values of the spin-orbit impurities concentrations. Purple plots are for purely scalar disorder ($\lambda =0$), while blue (green) plots for $\lambda = 0.2$ ($\lambda = 1.25$).
The magnetic field is given in the unit of $\tilde{B}_0 = e / (2 \pi \hbar v_\text{F}^2 \tau_0^2)$, where $\tau_0$ is the elastic scattering time in the absence of spin-orbit impurities and $v_\text{F}$ the Fermi velocity. Adapted from Figure~1 in Ref.~\cite{Adroguer15prb}, where $\delta\sigma(B) \equiv \Delta\sigma(B)$ as defined in Eq.~(\ref{magnetoconduc}).}
\label{fig:Qcorrec_HLN3D_alphaB}
\end{figure}
%------------------------------------------------------------

The HLN formula \cite{Hikami80ptp} describes the quantum correction to conductivity as a function of an applied magnetic field for non-relativistic electrons in presence of both scalar and spin-orbit impurities, where the only relevant parameter is the diffusion constant. In contrast, in our problem, each mode $\Gamma^\text{C}_{n,m}$ obeys a diffusion equation with a diffusion constant $D_i$ depending on $i = \vert n \vert + \vert m \vert$. Namely, these diffusion constants are $D_0 = v_\text{F}^2 \tau/(1 + \lambda^2/2)$, $D_1 = 2 v_\text{F}^2 \tau/\lambda $ and $D_2 = 2 v_\text{F}^2 \tau/ \lambda^2$. The magnetoconductivity corrections for massless Dirac fermions are shown in Figure~\ref{fig:Qcorrec_HLN3D_alphaB}. These results exhibit many differences from the HLN formula widely used to fit magnetotransport experiments of 3DTI surface states. The first is that the winding of the spin around the Fermi surface breaks the mirror symmetry around the $xy$-plane, so it is now possible to obtain a linear dependence of measurable quantities on $\lambda$, the strength of the spin-orbit disorder. Such a linear dependence is observed in the mean-free time $\tau$, the longitudinal Drude conductivity $\sigma_{xx}$, and the diffusion constant $D$. A second type of difference emerges from the anisotropy of the Green functions for massless Dirac model. This anisotropy together with the anisotropy coming from the spin-orbit impurities leads to 9 different Fourier modes in the Cooperon to the second order in  $\lambda$ as opposed to only one mode for the HLN model. Moreover, our expansion in the spin-orbit impurity strength shows more explicitly the fact that this calculation is perturbative in $\lambda$ and should be restricted to small values of the perturbative parameter as the odd powers of the series expansion contribute negatively to the conductivity.

In general symmetry terms, the massless Dirac fermion model stays in the symplectic class for all values of the impurity spin-orbit coupling, as the square of the time reversal operator $\Theta^2 = - \openone$. This explains why WAL is always observed for Dirac fermions. In the HLN formula, the introduction of the impurity spin-orbit coupling is responsible for a crossover from the orthogonal class $\Theta^2 = \openone$ when $\lambda=0$ to the symplectic ("pseudo-unitary"\footnote{Although TRS is preserved, the spin-orbit coupling only affects the $z$-component of the spin for a 2DEG, and the triplet state with no net magnetization along the $z$-axis is not suppressed. As a consequence, the singlet and triplet compensate each other, resulting in no correction to conductivity. This is similar to the unitary class where TRS is broken and all four states are suppressed.}) class when $\lambda \to \infty$ for 3DEGs (2DEGs).

\section{Weyl Semimetal Thin Films}
\label{Results}

% Highlight our contribution

\begin{figure}[h]
\centering
\includegraphics[width=0.3\columnwidth]{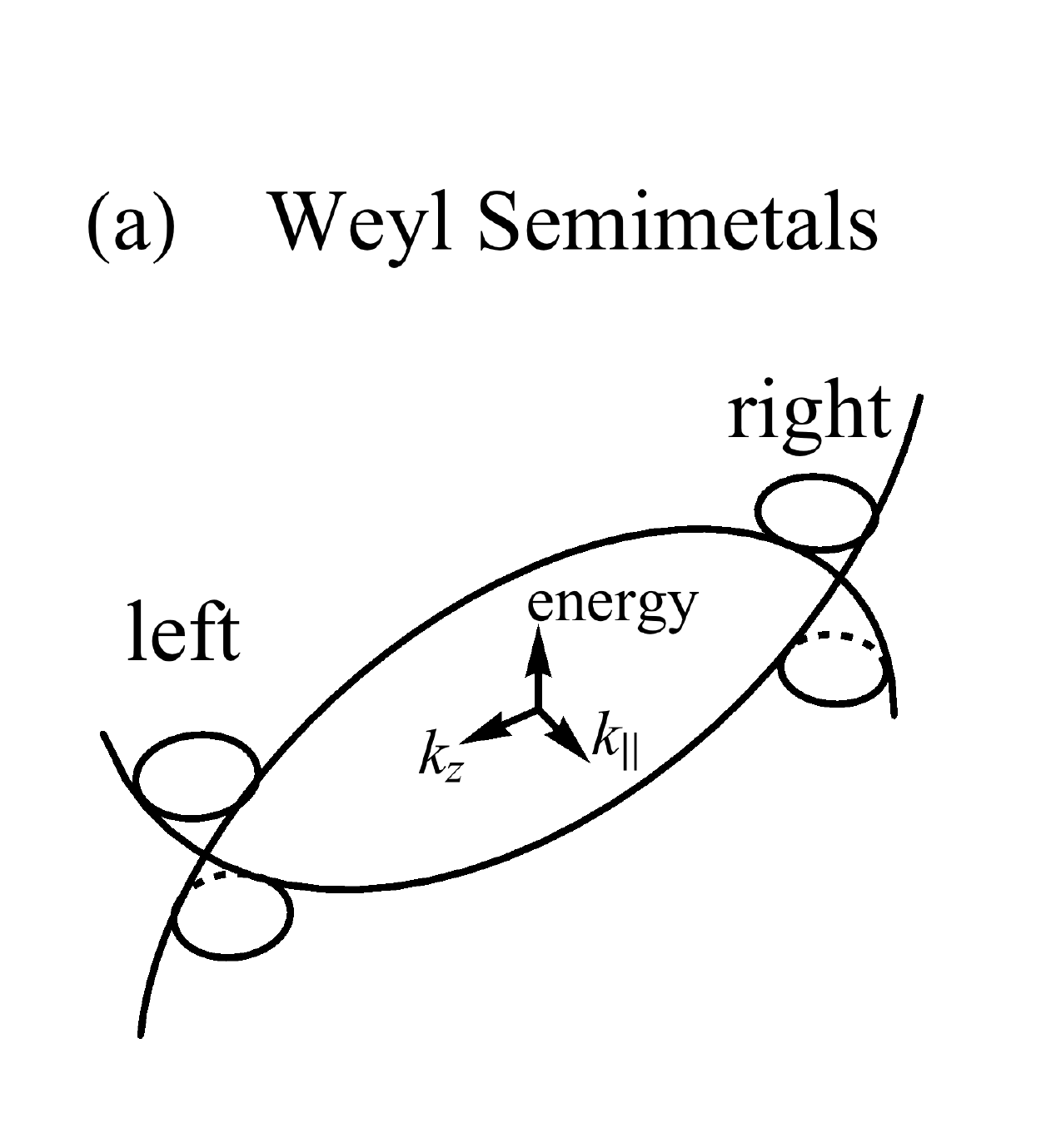}
\hspace{0.5cm}
\includegraphics[width=0.24\columnwidth]{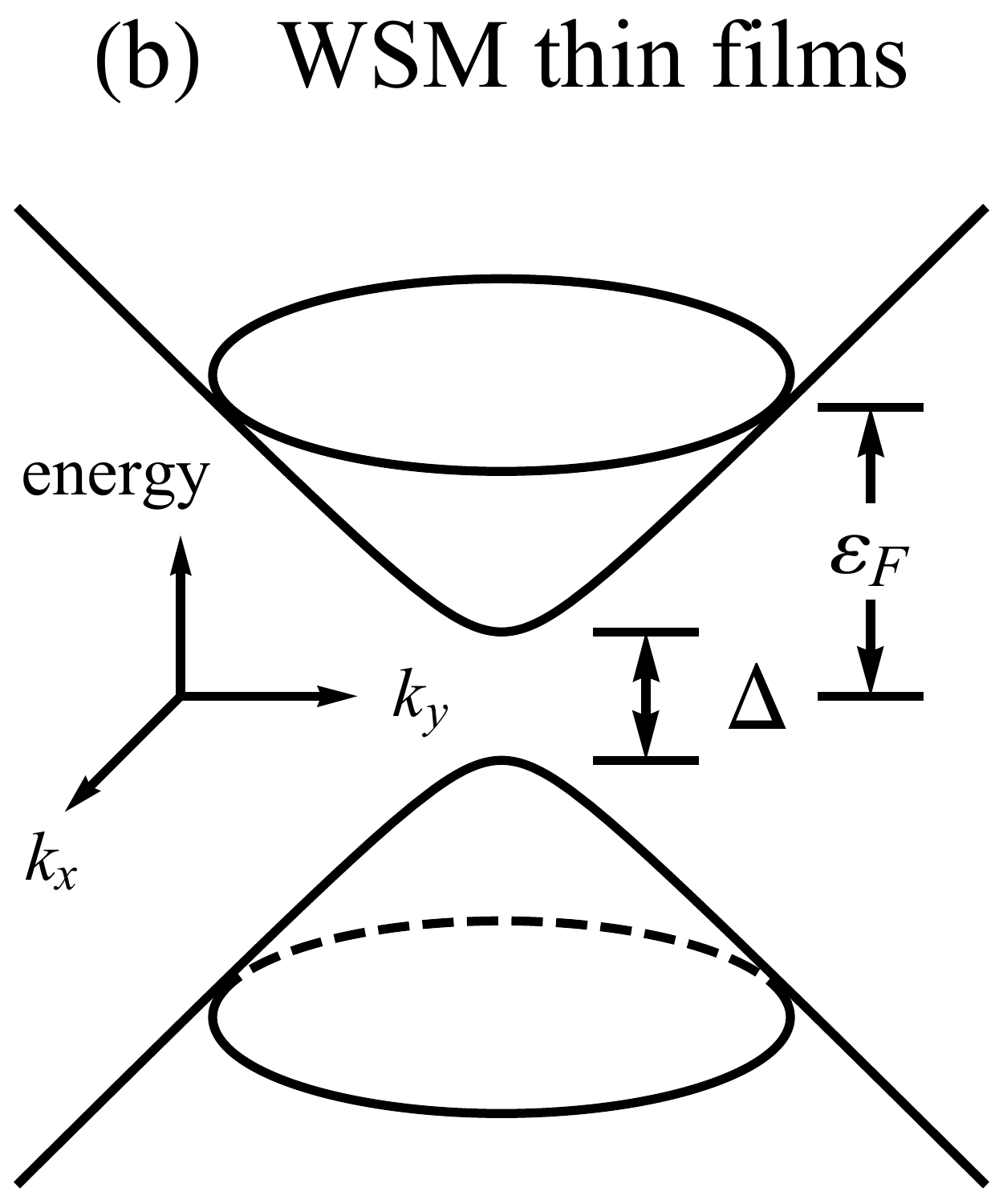}
\caption{\textbf{(a)} A minimal sketch of the energy dispersion of a Weyl semimetal. We have defined $k_{\parallel}^{2} = k_{x}^{2} + k_{y}^{2}$ while $(k_{x}, k_{y}, k_{z})$ represents the 3D wavevector.
$k_{z}$ points along the preferred direction.
The conductance and valence bands cross linearly at the Weyl nodes, i.e. the left and right cones, and the nodes appear in pairs with opposite chirality number. A Dirac node appears when two oppositely-chiral Weyl nodes emerge together. \textbf{(b)} A schematic picture of the band structure in WSM thin films, where $\Delta$ is the band gap and $\varepsilon_{\text{F}}$ is the Fermi energy.}
\label{Weyl_basic}
\end{figure}

In WSM thin films, the effective band Hamiltonian in the vicinity of a Weyl node is \cite{Lu2016,PhysRevB.72.045215}:
\begin{equation}\label{WeylHamEq}
H_{0\bm{k}} = A (\sigma_{x} k_{x} + \sigma_{y} k_{y}) + M \sigma_{z},
\end{equation}
where $A = \hbar v$ is a material-specific constant, $v$ is the effective velocity, $\bm{k} = (k_{x}, k_{y})$ is the in-plane wavevector measured from the Weyl node, and $M$ is the effective mass due to the quantum confinement. The band structure of bulk and thin-film WSM is sketched in Figure~\ref{Weyl_basic}. In (b) the gap is $ \Delta = 2 M$, which is the main distinguishing feature of WSM thin films as compared to topological insulators.

\begin{figure}[h]
\begin{center}
\includegraphics[width=0.5\columnwidth]{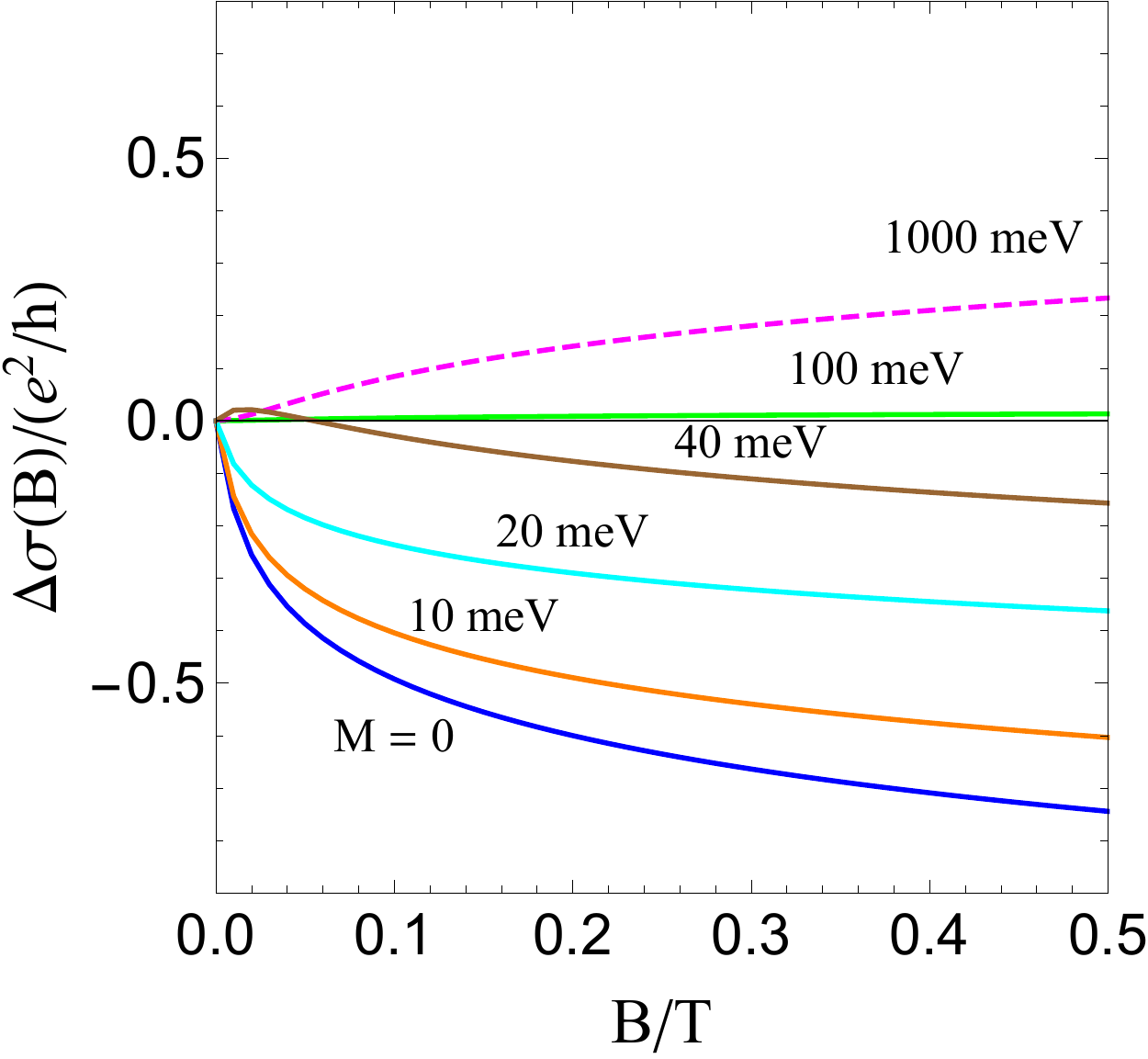}
\caption{The magnetoconductivity $\Delta \sigma (B)$ plots at $\lambda = 0.5$ for different masses $M$. The parameters used are $A = 300$ meV$\cdot$nm, $n_{e} = 0.01\text{ nm}^{-2},$ $n_{\text{i}} = 0.0001\text{ nm}^{-2}$ and $\ell_{\phi} = 500$ nm, according to Ref.~\cite{Lu11prl}.}
\label{eta1_Plot}
\end{center}
\end{figure}

Since the fermions are now massive there is a competition between WAL, driven by the spin-orbit term, and WL, driven by the mass term. The magnitude of the energy eigenvalues $\displaystyle \varepsilon_{k} = \sqrt{A^{2} k^{2} + M^{2}}$ tends to $Ak$ when the spin-orbit coupling dominates, which is identical to the case of topological insulators, and to $M$ when the mass dominates, which is identical to conventional electrons. The extrinsic spin-orbit coupling terms, which favor WAL, become very important when the band structure spin-orbit coupling and the mass are comparable in magnitude. To illustrate this, we first refer to Figure~\ref{eta1_Plot}, in which a strong suppression of the WL channel $(\alpha_{2})$ is seen due to the $\lambda$-linear terms, in particular in the green line. Moreover, the WAL channel number $(\alpha_{1})$ in the massless limit is exactly 1/2, which also satisfies the universal condition protected by time-reversal symmetry.

The zero-field quantum conductivity $\sigma^{\text{qi}} (0)$ can be used to locate the crossover between WAL and WL, and then separate out the regimes in which they occur, as shown in Figure~\ref{WALWLphase}. In Figure~\ref{WALWLphase} the $\lambda$-linear terms are taken into account from the start. A positive/negative $\sigma^{\text{qi}} (0)$ corresponds to WAL/WL. From Figure~\ref{WALWLphase}, in the absence of spin-orbit scattering, the WAL/WL transition occurs at $b \sim 0.3$. As the strength of the spin-orbit scattering is increased, the transition happens at larger $b$. The spin-orbit scattering pushes the WAL/WL boundary further into the WL regime, so the WAL regime becomes far broader. The behavior at large mass is similar to the 2D conventional electron gas case where the spin-orbit scattering drives the system from the WL to WAL regimes.

One interesting feature of Ref.~\cite{Tkachov11prb} is
the detailed study of the carrier density dependence.
Although those studies were only for the 2D TIs in Ref.~\cite{Tkachov11prb},
it motivated authors to investigate the carrier density dependence of the zero-field conductivity.
Note that a back gate is usually applied to adjust the carrier density,
whereas Hall measurements are performed to measure the density.
We introduce $\lambda = \lambda_{c} n_{e}$ since $\lambda \propto k_{\text{F}}^{2}$ gives a linear density dependence.

At zero temperature, the residual conductivity \cite{PhysRevLett.109.096801} is
$\sigma^{(0)} = \sigma^{\text{Dr}}_{xx} + \sigma^{\text{qi}}_{xx}$, and $\ell_{\phi}$ in Eq.~(\ref{zerofieldcondu}) should
be replaced by the sample size $L,$ because the former will be divergent if $T \to 0$.
By tuning the gate voltage, the carrier density dependence of $\sigma^{(0)}$
can be experimentally measured and the strength
of the spin-orbit scattering can be extracted from $\sigma^{(0)}$. This is one of the central arguments of this work.
To date in the literature only $\lambda^{2}$ terms in $\tau$ or $\tau_{\text{tr}}$ have been identified,
which arise when the spin-orbit scattering terms are averaged over directions in momentum space assuming a circular Fermi surface.
It contributes a negligible $n_{e}^{2}$ dependence to $\sigma^{(0)}$, which is effectively linear in $n_e$.
When the nontrivial linear-$\lambda$ term is taken into account in $\tau$ and $\tau_{\text{tr}}$,
which is caused by the non-commutativity of the band structure and the random spin-orbit impurity field, the WAL/WL correction has a more pronounced dependence on the carrier number density. This provides a new possibility to extract the spin-orbit scattering constant from the density dependence of $\sigma^{(0)}.$

Specifically, at small mass, the extrinsic spin-orbit scattering causes the momentum relaxation time to acquire a strong density dependence. This manifests itself in a flattening of the 2D WAL correction as a function of carrier number density. To extract the spin-orbit scattering strength $\lambda$ from $\sigma^{\text{qi}} (0)$ in the massless limit, it is possible to follow the fitting equation
\begin{equation}\label{fittingequation}
\sigma^{\text{qi}} (0) = a_{0} \ln n_{e} + b_{0} n_{e}.
\end{equation}
Here we assume zero temperature, so $\ell_{\phi}$ is replaced by the sample size $L$ as mentioned above. The extracted coefficient $b_{0}$ yields the spin-orbit scattering strength
$\lambda =  - b_{0} n_{e} / (3 e^{2} / 2 \pi h)$ when $\lambda \ll 1$. The short-range impurity potential that we have used in this article needs to be replaced by a long-range Coulomb potential (or a generic long-range potential). In the latter case, the Bethe-Salpeter equation will involve an additional angular integration over the impurity potential, in which case it is no longer solvable in closed form. One possible rough estimate for $\sigma^{\text{qi}} (0)$ can be obtained by retaining the form of Eq.~(\ref{zerofieldcondu}) while substituting the same $\tau$ for the long-range potential as was found above for the short-range case. In the long-range case, the normal impurity potential $\mathcal{U}$ for 2D massless Dirac fermions becomes $\mathcal{U}_{\text{long}} = Z e^{2} / [2 \varepsilon_{r} k_{\text{F}} \sin^{2} (\gamma / 2)],$\cite{PhysRevB.78.235417} where $\varepsilon_{r}$ is the material-specific dielectric constant. Thus, the long-range potential $\mathcal{U}_{\text{long}}$ will have the same $n_{e}$ dependence as the short-range case, and the fitting equation~(\ref{fittingequation}) will be still correct.

\begin{figure}[h]
\begin{center}
\includegraphics[width=0.5\columnwidth]{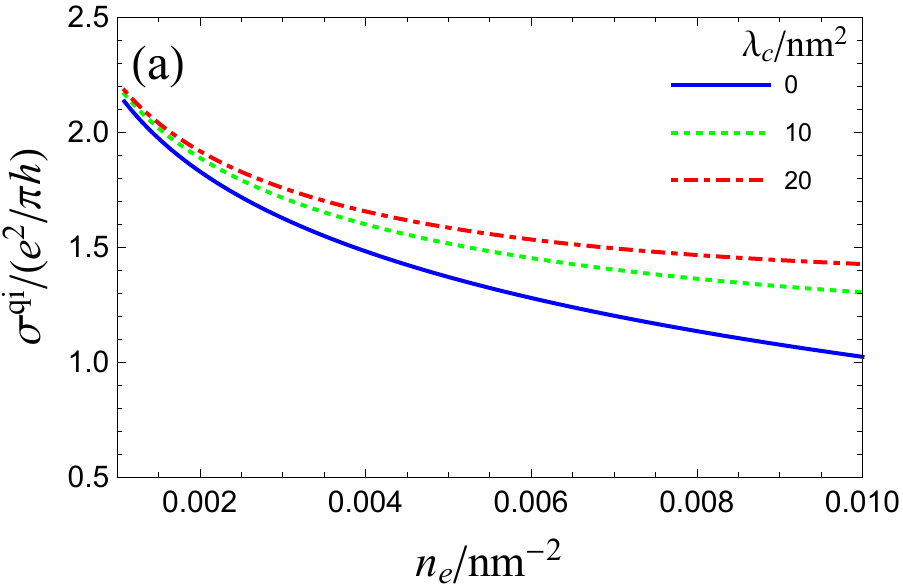}\\
\includegraphics[width=0.5\columnwidth]{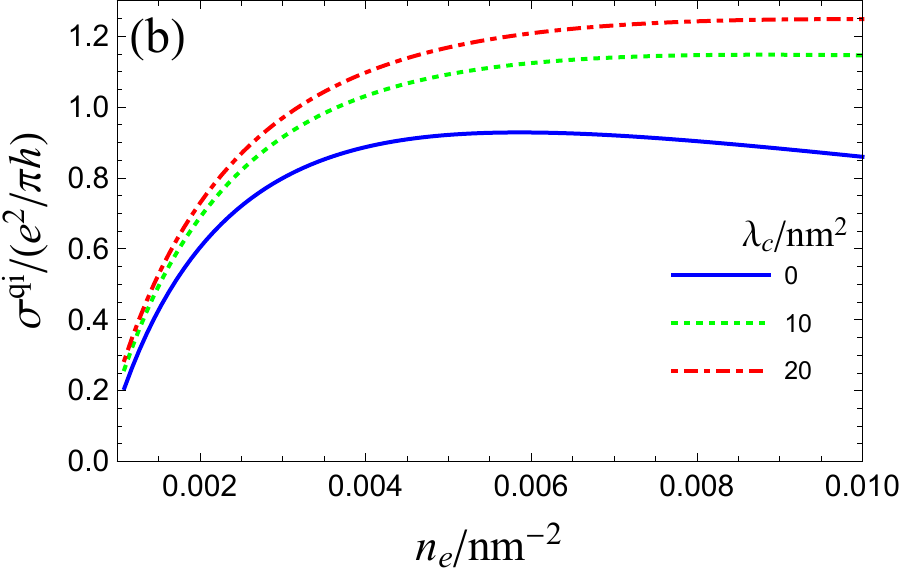}\\
\includegraphics[width=0.5\columnwidth]{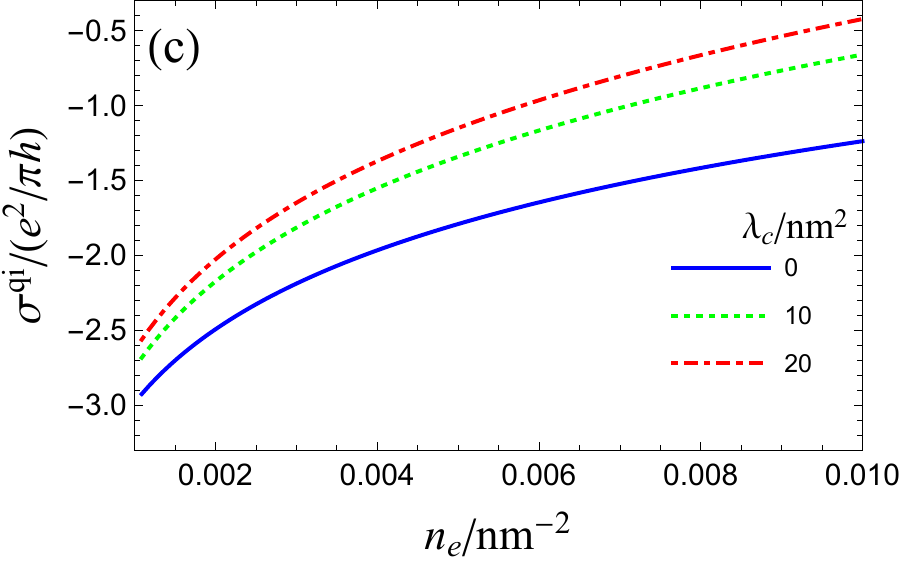}\\
\caption{\textbf{(a)}, \textbf{(b)} and \textbf{(c)} The carrier density dependence of the quantum-interference conductivity $\sigma^{\text{qi}}$ in the massless $(M = 0),$ small mass $( M = 10 \text{ meV}),$ and large mass $(M = 100 \text{ meV})$ cases. The parameters are the same as in Fig.~\ref{eta1_Plot} and $L = \ell_{\phi}$. Note that $\lambda = \lambda_{c} n_{e}$ and $ A k_{\text{F}} = 106$ meV for $n_{e} = 0.01\,\text{nm}^{-2}$.}
\label{Weyl_density_sigmqi}
\end{center}
\end{figure}

It is enlightening to plot the carrier density dependence of the quantum-interference part of the conductivity: this contribution can be singled out experimentally since it vanishes in an external magnetic field. Its carrier density dependence is plotted in Figure~\ref{Weyl_density_sigmqi} for three different masses : $M = 0,$ 10 meV, and 100 meV. In the massless limit, Figure~\ref{Weyl_density_sigmqi}(a), for $\lambda_c = 0$ the contribution $\sigma^{\text{qi}}_{xx}$ follows the logarithmic dependence expected of a 2DEG. As $\lambda_c$ increases this logarithm becomes nearly flat at large enough densities. This altered density dependence arises from the linear terms in the scattering and transport times. In the small mass limit, Figure~\ref{Weyl_density_sigmqi}(b), a sharp suppression of the conductivity at the small density is displayed, because the effect of the mass term becomes more significant when the carrier density decreases and the WAL channel is suppressed. In the large mass limit, see Figure~\ref{Weyl_density_sigmqi}(c), a negative conductivity correction (WL) is expected.

Indeed the presence of the mass is a crucial difference between Weyl semimetal films and topological insulators, and may offer the possibility of distinguishing between TI states and WSM states in certain parameter regimes. In Weyl semimetals the sign of the correction depends on the quasiparticle mass: at small mass it is identical to topological insulators (weak antilocalization), whereas at large mass it is similar to ordinary massive fermions (weak localization).

Extrinsic spin-orbit scattering affects the transition from WL to WAL as a function of the mass. This may be seen in Figure~\ref{WALWLphase}. At small and large values of the mass, as expected, the extrinsic spin-orbit scattering has little effect on the WL/WAL correction. At large mass, moreover, the extrinsic spin-orbit scattering also plays a negligible role in the momentum relaxation time. At certain intermediate values of the mass, however, the extrinsic spin-orbit scattering is critical in determining whether the system experiences WL or WAL.
It is rough that the WL/WAL transition line in Figure~\ref{WALWLphase} is $b \sim 0.3 + 0.8 \cdot \lambda$, which exists when $n_{e} \in [0.01, 0.02] \, \text{nm}^{-2}$ and $\ell_{\phi} \in [200,500] \, \text{nm}$. This fitting equation can provide a semi-empirical formula to extract the $\lambda$.

\section{Other topological materials}

Aside from topological insulators and Weyl semimetal films, another series of topological materials that have received considerable attention are transition metal dichalcogenides such as MoS$_2$. These materials exhibit some qualitative differences as compared to those studied so far, not least through the presence of a lattice pseudospin analogous to that found in graphene. The Hamiltonian for transition metal dichalcogenides is given in \cite{Dixiao_MoS2_2012} as
\begin{equation}
H_{tmd} = A (v \sigma_x k_x + \sigma_y k_y) + \frac{\Delta}{2} \, \tau_z + v \varepsilon_s \bigg(\frac{\openone - \tau_z}{2} \bigg) \, \sigma_z.
\end{equation}
Here $v = \pm$ represents the valley index, $\Delta$ is an energy gap analogous to the mass term in WSM films, $\varepsilon_s$ is the energy gap at the valence band top induced by spin-orbit coupling, while the Pauli matrix $\tau_z$ represents the sublattice pseudospin. Upon inspection of this Hamiltonian it is immediately obvious that the physics of transition metal dichalcogenides is governed by the interplay of the valley, spin and pseudospin degrees of freedom, in close analogy with graphene but with the proviso that the spin-orbit coupling is extremely strong. It is well known that in graphene monolayers and bilayers the interplay of valley and pseudospin physics strongly affects the physics of weak localization/antilocalization, in a manner that is not captured in this work \cite{Kechedzhi07eurphysjspecialtopgics}. For this reason we have not included graphene in this study, and we do not include transition metal dichalcogenides, where determination of the exact functional form of the extrinsic spin-orbit coupling is also nontrivial.

However, on a qualitative level it is natural to expect that extrinsic spin-orbit coupling terms will play an important role in WL/WAL in transition metal dichalcogenides, and due to the same non-commutativity of spin matrices we have identified we expect terms linear in the extrinsic spin-orbit coupling to be present in WL/WAL in these materials. They may give rise to a similar density dependence in the Bloch and transport times, and affect the WL/WAL transition. 

\section{Conclusions}\label{Conclusions}

The magnetoconductivity corrections for topological semimetals in the presence of scalar and spin-orbit impurity disorder exhibit profound differences from the HLN model for conventional electrons with parabolic dispersion. In both cases the diffusion constant and the longitudinal conductivity are renormalized to the first order in the spin-orbit scattering strength. In topological insulators WAL is expected to occur in the presence as well as in the absence of spin-orbit impurity scattering, and the correction has a linear dependence on the carrier density. In WSM thin films terms linear in the extrinsic spin-orbit scattering strength play an important role in determining whether the system experiences WL or WAL and strongly affect the density dependence of the WAL correction in the massless limit. These observations are directly relevant to the analysis of transport experiments in topological semimetals, which continues to rely heavily on the HLN formula. 

\acknowledgments{We thank Hai-Zhou Lu and Shun-Qing Shen for a series of illuminating discussions. This research was supported by the Australian Research Council Centre of Excellence in Future Low-Energy Electronics Technologies (project number CE170100039) and funded by the Australian Government. E. M. H. acknowledges financial support by the German Science Foundation (DFG) via SFB 1170 "ToCoTronics" and the ENB Graduate School on Topological Insulators.}
 
%\externalbibliography{yes}
%\bibliography{Materials}

\end{document}